\documentclass[epj,final]{svjour}
\usepackage{graphics}
\input{epsf}
\begin{document}
\title{Effects of single-qubit quantum noise on entanglement purification}
\author{Giuliano Benenti\inst{1,2}, Sara Felloni\inst{3},
and Giuliano Strini\inst{4}}
\institute{Center for Nonlinear and Complex Systems,
Universit\`a degli Studi dell'Insubria,
Via Valleggio 11, 22100 Como, Italy
\and
Istituto Nazionale per la Fisica della Materia, Unit\`a di Como and
Istituto Nazionale di Fisica Nucleare, Sezione di Milano
\and
Dipartimento di Matematica, Universit\`a degli Studi di Milano,
via Saldini 50, 20133 Milano, Italy
\and
Dipartimento di Fisica, Universit\`a degli Studi di Milano,
via Celoria 16, 20133 Milano, Italy}
\titlerunning{Effects of single-qubit quantum noise on entanglement 
purification}
\authorrunning{G. Benenti, S. Felloni, and G. Strini}
\date{Received: September 19, 2005}
\abstract{We study the stability under quantum noise effects 
of the quantum privacy amplification protocol for 
the purification of entanglement in quantum cryptography. 
We assume that the E91 protocol is used by two communicating
parties (Alice and Bob) and that the eavesdropper Eve uses the 
isotropic Bu\v{z}ek-Hillery quantum copying machine to extract 
information. 
Entanglement purification is then operated by Alice and Bob
by means of the quantum privacy amplification protocol and we 
present a systematic numerical study of the impact of all possible
single-qubit noise channels on this protocol. We find that 
both the qualitative behavior of the fidelity of the purified
state as a function of the number of purification steps and
the maximum level of noise that can be tolerated by the 
protocol strongly depend on the specific noise channel. 
These results provide valuable information for experimental
implementations of the quantum privacy amplification protocol.}

\maketitle

\section{Introduction} 
\label{sec:intro}

A central problem of quantum communication is how to reliably
transmit quantum information through a noisy quantum channel.
The carriers of information (the qubits) unavoidably interact
with the external world, leading to phenomena such as decoherence
and absorption. In particular, if a member of a maximally 
entangled EPR (Einstein-Podolsky-Rosen) pair is transmitted 
from a sender (known as Alice) 
to a receiver (Bob) through a quantum channel, then noise in 
the channel can degrade the amount of entanglement of the pair.
This problem is of primary importance for entanglement-based 
quantum cryptography. Indeed, in the idealized E91 protocol 
\cite{E91} Alice and Bob share a large number of maximally 
entangled states. 
Entanglement purification techniques exist \cite{Bennett,Bennett2}.
In particular, they have been applied to quantum cryptography:
in Ref.~\cite{DEJMPS} a quantum privacy amplification (QPA) 
iterative protocol was proposed, that eliminates 
entanglement with an eavesdropper by creating a small
number of nearly perfect (pure) EPR states out of a large number 
of partially entangled states.
This protocol is based on the so-called LOCC,
that is on local quantum operations
(quantum gates and measurements performed by Alice and Bob 
on their own qubits), supplemented by classical communication. 

Under realistic conditions, the quantum operations themselves 
are unavoidably affected by errors and introduce a certain 
amount of noise. A first study of the impact
of these errors on the QPA protocol was made in Ref.~\cite{briegel}
and conditions for the security of QPA were found. However, 
the noise model considered in \cite{briegel} was not the most 
general one. In particular, error channels like the amplitude 
damping or thermal excitations were not considered. 

Studies of the impact of noise on the stability of quantum computation 
and communication are of primary importance for the practical 
implementation of quantum information protocols. 
In this paper, for the first time {\it all} single-qubit quantum 
noise channels are studied and compared and their different impact 
on the quantum privacy amplification protocol is elucidated.
Errors acting on single qubits are described most conveniently using 
the Bloch sphere picture: Quantum noise acting on a single qubit is 
described by 12 parameters, associated to rotations, deformations
and displacements of the Bloch sphere. We study in detail the 
effects of these different errors and show that they impact {\it very
differently} on the QPA algorithm. In particular, errors giving 
a displacement of the Bloch sphere are very dangerous.
These results provide valuable information for experimentalists:
indeed, knowing what are the most dangerous 
noise channels is useful to address experiments towards 
implementations for which these channels have negligible impact.

The paper is organized as follows. The eavesdropper's attack 
strategy is described in Sec.~\ref{sec:Eve}.
Here we assume that the eavesdropper Eve
attacks the qubits sent by Alice to Bob by means of the quantum 
copying machine of Bu\v{z}ek and Hillery \cite{buzekhillery}.
As a result, Alice and Bob share partially entangled pairs.
Each pair is now entangled with the environment (Eve's qubits)
and described by a density operator. 
The QPA protocol, reviewed in Sec.~\ref{sec:QPA}, can be used to 
purify entanglement and, as a consequence, reduce the entanglement 
with any outside system to arbitrarily low values 
(a maximally entangled EPR pair is a pure state automatically 
deentangled from the outside world). 
We then consider the effects of noise acting on the purification 
protocol. The most general single-qubit noise channels are discussed in 
Sec.~\ref{sec:Blocherrors}. We model each noise channel by means of 
equivalent quantum circuits, from which the usual Kraus representation
and the transformation (rotation, translation or displacement) 
of the Bloch sphere coordinates can be derived. 
The impact of these errors on the entanglement purification is     
discussed in Sec.~\ref{sec:noisyQPA}.
Finally, in Sec.~\ref{sec:conc} we present our conclusions.

\section{Eavesdropping} 
\label{sec:Eve}

We assume that Alice has at her disposal a source of EPR pairs
and sends a member of each pair to Bob.
The eavesdropper Eve wants, on one hand, to find out as much information
as possible on the transmitted qubits and, on the other hand, make his
intrusion as unknown as possible to Alice and Bob. 
Isotropic cloning by means of the
Buzek-Hillery machine \cite{buzekhillery} is the most natural way 
to meet these two requirements. We also note that isotropy is 
necessary only in the case in which Alice and Bob use a six-state
protocol, that is, the measurements are performed along the 
$x$, $y$ and $z$ axis of the Bloch sphere. 
The isotropy condition may be relaxed when Alice and Bob use a 
four-state protocol: they measure only along $x$ and $z$  
and Eve knows what are the measurement axes. 
In this case, it would be sufficient for Eve to send Bob qubits that reproduce
as faithfully as possible the $x$ and $z$ coordinates, but with no constraints
about $y$. We have also studied this case (non isotropic cloning) but
not reported it on the paper for the sake of simplicity.

\begin{figure}
\centerline{\epsfxsize=8.cm\epsffile{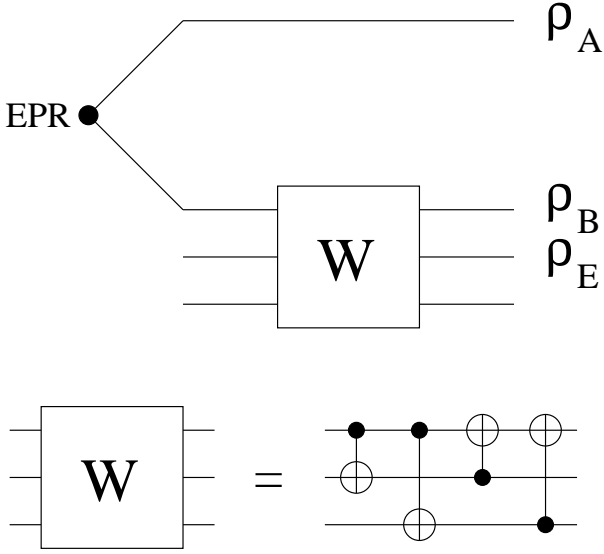}}
\caption{Top: quantum circuit representing the intrusion 
(by means of the Bu\v{z}ek-Hillery copying machine) of the 
eavesdropper Eve in the E91 protocol. The density matrices
$\rho_A$, $\rho_B$ and $\rho_E$ represent the states of Alice's qubit,
Bob's qubit and Eve's qubit after tracing over all other qubits. 
Bottom: decomposition of the unitary transformation $W$ 
in four CNOT gates. By definition, CNOT$|x\rangle|y\rangle=
|x\rangle|y\oplus x\rangle$, with $x,y=0,1$ and $\oplus$ indicating
addition modulo 2. 
The first ($x$) qubit in the CNOT gate acts as a control 
(full circle in the figure) and the second ($y$) as a
target qubit ($\oplus$ symbol).
Here and in the following circuits, any sequence of
logic gates must be read from the left (input) to the right (output). 
From bottom to top, qubits run from the least significant to the most
significant.}
\label{figBHcrypto}
\end{figure}

\begin{figure}
\centerline{\epsfxsize=8.cm\epsffile{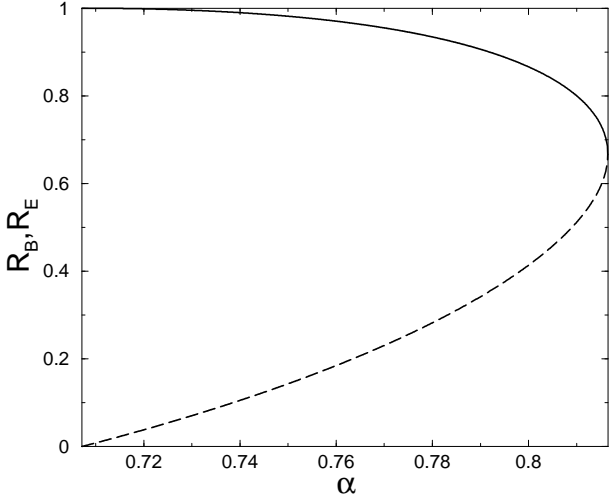}}
\caption{Ratios $R_B$ (solid line) and $R_E$ (dashed line) for
the isotropic Bu\v{z}ek-Hillery copying machine versus 
the parameter $\alpha$.}
\label{bobeveratios}
\end{figure}

In the following we assume that, 
as shown in Fig.~\ref{figBHcrypto}, Eve attacks the qubits sent by 
Alice using the Bu\v{z}ek-Hillery machine \cite{buzekhillery}.
The two bottom qubits in Fig.~\ref{figBHcrypto} are prepared 
by Eve in the state
\begin{equation}
|\Phi\rangle = \alpha |00\rangle+\beta|01\rangle+\gamma|10\rangle+
\delta|11\rangle
\label{betgamdel}
\end{equation}
and we assume that $\alpha,\beta,\gamma,\delta$ are
real parameters. Let us call $\rho_B$ and $\rho_E$ the density matrices
describing the final states of Bob's qubit and Eve's qubit.
As we have said, we assume isotropy, that is, if we call
$(x,y,z)$ the coordinates of the qubit sent from Alice to
Bob before eavesdropping, then the Bloch sphere coordinates
$(x_B,y_B,z_B)$ and $(x_E,y_E,z_E)$ associated to
$\rho_B$ and $\rho_E$ are such that $x_B/x=y_B/y=z_B/z\equiv R_B$ 
and $x_E/x=y_E/y=z_E/z\equiv R_E$.
As shown in Appendix~\ref{app:isotropiccloning},
these conditions are fulfilled for
\begin{equation}
\beta=\frac{\alpha}{2}-\sqrt{\frac{1}{2}-\frac{3}{4}\alpha^2},\quad
\gamma=0,\quad
\delta=\frac{\alpha}{2}+\sqrt{\frac{1}{2}-\frac{3}{4}\alpha^2}.
\label{betadeltabh}
\end{equation}
It can be checked by direct computation 
(see again Appendix~\ref{app:isotropiccloning}) that
in this case $(x_B,y_B,z_B)=2\alpha\delta(x,y,z)$
and $(x_E,y_E,z_E)=2\alpha\beta(x,y,z)$. Since the Bloch
sphere coordinates must be real and nonnegative, we obtain
$\frac{1}{\sqrt{2}}\le \alpha \le \frac{2}{\sqrt{6}}$.
The ratios $R_B\equiv x_B/x=y_B/y=z_B/z$ and
$R_E\equiv x_E/x=y_E/y=z_E/z$ are shown in Fig.~\ref{bobeveratios}.
It can be seen that the two limiting cases $\alpha=\frac{1}{\sqrt{2}}$
and $\alpha=\frac{2}{\sqrt{6}}$ correspond
to no intrusion ($x_B=x,y_B=y,z_B=z$)
and maximum intrusion ($x_E=x_B,y_E=y_B,z_E=z_B$), respectively.
In the first case, the qubit sent from Alice to Bob is not attacked. 
In the latter case Eve makes two imperfect 
identical copies of the original qubit
(symmetric Bu\v{z}ek-Hillery machine), that is $\rho_E=\rho_B$: in
this way Eve both optimizes the information obtained about the 
transmitted state and minimizes the modification of the qubit 
received by Bob.
The degree of Eve's intrusion is therefore
conveniently measured by the intrusion parameter
\begin{equation}
f_\alpha=\frac{\alpha-\frac{1}{\sqrt{2}}}{\frac{2}{\sqrt{6}}-
\frac{1}{\sqrt{2}}},
\end{equation}
with $0\le f_\alpha \le 1$.

\section{Quantum privacy amplification}
\label{sec:QPA}

\begin{figure}
\centerline{\epsfxsize=8.cm\epsffile{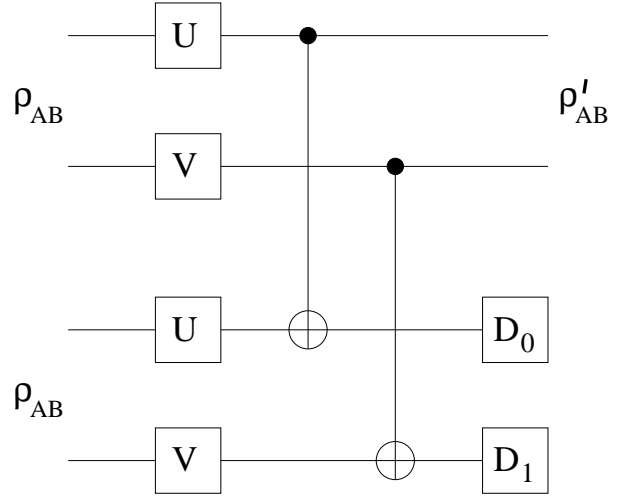}}
\caption{Schematic drawing of the QPA entanglement purification
scheme. Note that the density matrix $\rho_{AB}^\prime$ describes
the two top qubits only when the detectors $D_0$ and $D_1$ give
the same outcome.}
\label{figDEJMPS}
\end{figure}

We assume that Alice and Bob purify entanglement by means of the
QPA protocol \cite{DEJMPS}. This is an iterative procedure, which
we briefly review in what follows. 
At each iteration, the EPR pairs are combined in groups of two.
The following steps are then taken for each group
(see Fig.~\ref{figDEJMPS}):
\begin{itemize}
\item
Alice applies to her qubits a $\frac{\pi}{2}$ rotation about the
$x$-axis of the Bloch sphere, described by the unitary matrix
\begin{equation}
U=
R_x\left(\frac{\pi}{2}\right)=
\frac{1}{\sqrt{2}}\left[
\begin{array}{cc}
1 & -i \\
-i & 1
\end{array}
\right].
\end{equation}
\item
Bob applies to his qubits the inverse operation
\begin{equation}
V=
U^{-1}=
R_x\left(-\frac{\pi}{2}\right)=
\frac{1}{\sqrt{2}}\left[
\begin{array}{cc}
1 & i \\
i & 1
\end{array}
\right].
\end{equation}
\item
Both Alice and Bob perform a CNOT gate 
(defined in the caption of Fig.~\ref{figBHcrypto}) using their members
of the two EPR pairs.
\item
They measure the polarizations $\sigma_z$ of the two target qubits.
\item
Alice and Bob compare the measurement outcomes by means of a public
classical communication channel. If the outcomes coincide,
the control pair is kept for the next iteration and the target pair
discarded. Otherwise, both pairs are discarded.
\end{itemize}

\begin{figure}
\centerline{\epsfxsize=8.cm\epsffile{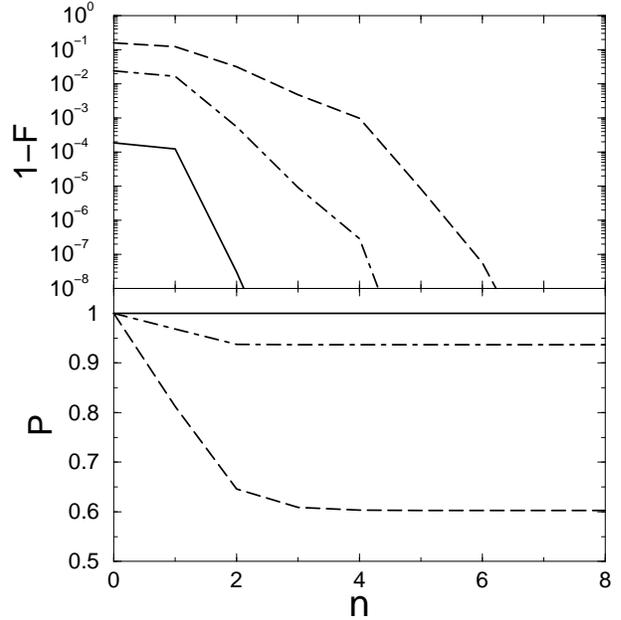}}
\caption{Deviation $1-F$ of the fidelity $F$ from the ideal
case $F=1$ (top) and survival probability $P$ (bottom) 
as a function of the number of iterations $n$ of map
(\ref{DEJMPSmap}). The different curves correspond
to the intrusion parameter
$f_\alpha=0.95$ (dashed line), $0.5$ (dot-dashed line)
and $0.05$ (solid line).}
\label{DEJMPSpure}
\end{figure}

In order to illustrate the working of the 
QPA procedure, let us consider the special
case in which the initial mixed pairs are described by
the density matrix $\rho_{AB}$ obtained from the ideal EPR
state $|\phi^+\rangle=\frac{1}{\sqrt{2}}(|00\rangle+|11\rangle)$
after application of the Bu\v{z}ek-Hillery copying machine with
intrusion parameter $f_\alpha$. After application of the unitary
transformation $W$ in Fig.~\ref{figBHcrypto} the overall state of 
the four-qubit system becomes
\begin{equation}
\begin{array}{l}
\frac{1}{\sqrt{2}}(\alpha|0000\rangle+\beta|0101\rangle+\gamma|0110\rangle
+\delta|0011\rangle
\\
+\alpha|1111\rangle+\beta|1010\rangle+\gamma|1001\rangle
+\delta|1100\rangle).
\end{array}
\end{equation}
After tracing over Eve's two qubits, we obtain
\begin{equation}
\rho_{AB}=\frac{1}{2}
\left[
\begin{array}{cccc}
\alpha^2+\delta^2 & 0 & 0 & 2\alpha\delta \\
0 & \beta^2+\gamma^2 & 2\beta\gamma & 0 \\
0 & 2\beta\gamma & \beta^2+\gamma^2 & 0 \\
2\alpha\delta & 0 & 0 & \alpha^2+\delta^2
\end{array}
\right].
\end{equation}
We note that this state is diagonal in the so-called Bell basis
$\{|\phi^{\pm}\rangle=\frac{1}{\sqrt{2}}(|00\rangle\pm |11\rangle),
|\psi^{\pm}\rangle=\frac{1}{\sqrt{2}}(|01\rangle\pm |10\rangle)\}$.
Indeed, we have
\begin{equation}
\begin{array}{c}
\rho_{AB}= A |\phi^+ \rangle\langle \phi^+ |+
B |\phi^- \rangle\langle \phi^- |\\\\
+C |\psi^+ \rangle\langle \psi^+ |+
D |\psi^- \rangle\langle \psi^- |,
\end{array}
\label{belldiagonal}
\end{equation}
where $A=\frac{1}{2}(\alpha+\delta)^2$,
$B=\frac{1}{2}(\alpha-\delta)^2$,
$C=\frac{1}{2}(\beta+\gamma)^2$ and
$D=\frac{1}{2}(\beta-\gamma)^2$.
The quantum circuit in Fig.~\ref{figDEJMPS} maps the state
$\rho_{AB}$ of the control pair, in the case in which it is not discarded,
onto another state $\rho_{AB}^\prime$ diagonal in the Bell basis.
Namely, $\rho_{AB}^\prime$ can be expressed in the form (\ref{belldiagonal}),
provided that new coefficients $(A^\prime,B^\prime,C^\prime,D^\prime)$
are used instead of $(A,B,C,D)$:
\begin{equation}
\begin{array}{c}
A^\prime=\frac{A^2+D^2}{N},\quad
B^\prime=\frac{2AD}{N},\\\\
C^\prime=\frac{B^2+C^2}{N},\quad
D^\prime=\frac{2BC}{N},
\end{array}
\label{DEJMPSmap}
\end{equation}
where $N=(A+D)^2+(B+C)^2$ is the probability that Alice and Bob
obtain coinciding outcomes in the measurement of the target qubits.
Note that map (\ref{DEJMPSmap}) is nonlinear as a consequence of
the strong nonlinearity of the measurement process.
The fidelity after the purification procedure is given by
\begin{equation}
F = \langle \phi^+ | \rho_{AB}^\prime | \phi^+ \rangle
\label{fido}
\end{equation}
(note that $F=A^\prime$).
This quantity measures the probability that the control qubits would
pass a test for being in the state $|\phi^+\rangle$.
Map (\ref{DEJMPSmap}) can be iterated and we wish to drive the fidelity
to one. It is possible to prove
\cite{macchiavello98} that this map converges to the target point
$A=1,B=C=D=0$ for all initial states (\ref{belldiagonal}) with
$A>\frac{1}{2}$.
This means that, when this condition is satisfied and
a sufficiently large number of initial pairs is available,
Alice and Bob can distill asymptotically pure EPR pairs.
Note that the quantum privacy amplification procedure is rather wasteful,
since at least half of the pairs (the target pairs) are lost at every
iteration. This means that to extract one pair close to
the ideal EPR state after $n$ steps we need at least
$2^n$ mixed pairs at the beginning. 
However, this number can be significantly larger, since pairs 
must be discarded when Alice and Bob obtain different 
measurement outcomes. 
We therefore compute the survival probability $P(n)$,
measuring the probability that a $n$-step QPA protocol
is successful. More precisely, if $p_i$ is the probability
that Alice and Bob obtain coinciding outcomes at step $i$, 
we have 
\begin{equation}
P(n)=\prod_{i=1}^n p_i.
\label{survp}
\end{equation}
The efficiency $\xi(n)$ of the algorithm is given by the 
number of obtained pure EPR pairs divided by the number 
of initial impure EPR pairs. We have 
\begin{equation}
\xi(n)=\frac{P(n)}{2^n}.
\end{equation} 
Both the fidelity and the survival probability are
shown in Fig.~\ref{DEJMPSpure}. The different curves of this
figure correspond to values of the intrusion parameter from
$f_\alpha=0.05$ (weak intrusion) to $f_\alpha=0.95$ (strong
intrusion). 
It can be seen that the convergence of the QPA
protocol is fast: the fidelity deviates from the
ideal case $F=1$ by less than $10^{-7}$ in no more than $n=6$
map iterations. Moreover, the survival probability is 
quite high: it saturates to  
$P_\infty\equiv \lim_{n\to\infty} P(n)= 0.60$ for $f_\alpha=0.95$, 
$P_\infty=0.94$ for $f_\alpha=0.5$ and 
$P_\infty=0.9995$ for $f_\alpha=0.05$.

\section{Single qubit errors}
\label{sec:Blocherrors}

In any realistic implementation of the QPA protocol, errors
acting on the purification operations are unavoidable. 
For the sake of simplicity we limit ourselves to consider
errors affecting only a single qubit. 
Nevertheless, we would like to stress that 
a complete treatment of the effects of all possible 
single-qubit noise channels on the QPA algorithm is 
provided in this paper. 

We need 12 parameters to characterize a generic
quantum noise operation acting on a single qubit \cite{chuang}.
Each parameter describes a particular noise channel
(like bit flip, phase flip, amplitude damping,...)
and can be most conveniently visualized as
associated to rotations, deformations and displacements 
of the Bloch sphere. 
In the following, we provide, for each noise channel,

(i) the Kraus representation, 

(ii) the transformation of the Bloch
sphere coordinates,  

(iii) an equivalent quantum circuit leading 
to a unitary representation in an extended Hilbert space.
A great advantage of these equivalent quantum circuits 
is that the evolution of the reduced density matrix 
describing the single-qubit system is automatically guaranteed
to be completely positive.

\begin{itemize}

\item
{\it Rotations of the Bloch sphere} - Rotations through an angle
$\theta$ about an arbitrary axis directed along the 
unit vector ${\bf n}$ are given by the operator \cite{qcbook} 
\begin{equation}
R_n(\theta)=\left(\cos\frac{\theta}{2}\right)I-
i\left(\sin\frac{\theta}{2}\right){\bf n}\cdot 
\mbox{\boldmath$\sigma$},
\end{equation}
where 
$\mbox{\boldmath$\sigma$}=(\sigma_x,\sigma_y,\sigma_z)$, 
$\sigma_x,\,\sigma_y$ and  $\sigma_z$ being the Pauli matrices.
The quantum circuit representing rotations is shown in Fig.~\ref{rotation}.
Any generic rotation can be obtained by composing 
rotations about the axes $x$, $y$ and $z$. 
Let us write as an example the transformation of the Bloch sphere coordinates
associated to a rotation through an angle $\theta$ about the $z$-axis:
\begin{equation}
\left\{
\begin{array}{l}
x^\prime=(\cos\theta)x - (\sin \theta)y,\\
y^\prime=(\sin\theta)x + (\cos \theta)y,\\
z^\prime=z
\end{array}
\right.
\end{equation}

\begin{figure}
\centerline{\epsfxsize=6.cm\epsffile{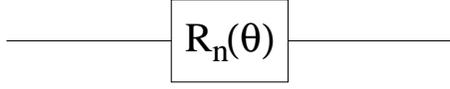}}
\caption{Quantum circuit representing a rotation through an angle $\theta$
about the ${\bf n}$-axis.}
\label{rotation}
\end{figure}

\item
{\it Deformations of the Bloch sphere} - 
The well known bit flip, phase flip and bit-phase flip channels
correspond to deformations of the Bloch sphere into an ellipsoid.
An equivalent quantum circuit implementing the bit-flip channel is 
shown in Fig.~\ref{bitflipcircuit}. Note that a single auxiliary qubit, 
initially prepared in the state $|\psi\rangle=
\cos\frac{\theta}{2}|0\rangle+\sin\frac{\theta}{2}|1\rangle$
(with $0\le\theta\le \pi$)
is sufficient to obtain a unitary representation of this noise 
channel. 
The corresponding Kraus representation is defined by the 
Kraus operators
\begin{equation}
F_0=\left(\cos\frac{\theta}{2}\right) I, \;\;\;
F_1=\left(\sin\frac{\theta}{2}\right) \sigma_x.
\end{equation}
The quantum operation 
\begin{equation}
\rho^\prime = \sum_k F_k \rho F_k^\dagger,\;\;\;
(\sum_k F_k^\dagger F_k=I),
\label{kraus}
\end{equation}
maps the Bloch sphere into an ellipsoid with $x$ as symmetry axis:
\begin{equation}
\left\{
\begin{array}{l}
x^\prime=x,\\
y^\prime=(\cos \theta) y,\\
z^\prime=(\cos \theta) z,
\end{array}
\right.
\end{equation}
The phase flip and bit-phase flip channels are obtained from 
quantum circuits analogous to Fig.~\ref{bitflipcircuit}, 
after substitution of $\sigma_x$ with $\sigma_z$ and 
$\sigma_y$, respectively. 
In the phase flip channel the Bloch sphere is mapped into an 
ellipsoid with $z$ as symmetry axis,
while in the bit-phase flip channel the symmetry axis is $y$.

\begin{figure}
\centerline{\epsfxsize=6.cm\epsffile{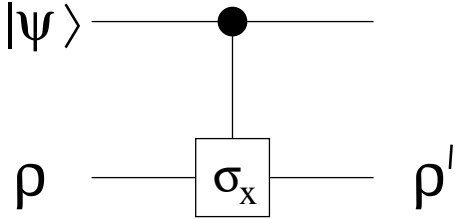}}
\caption{Quantum circuit implementing the bit flip channel.}
\label{bitflipcircuit}
\end{figure}

\item
{\it Displacements of the Bloch sphere} - 
A displacement of the center of the Bloch sphere must go 
with a deformation of the sphere. 
This is necessary if we want that $\rho^\prime$ still represents 
a density matrix: the Bloch radius ${\bf r}$ 
associated to any density matrix must have length $r$ such that
$0\le r\le 1$. 
This condition can be fulfilled as follows. Let us consider, for
instance, a displacement of the center of the Bloch sphere 
along the $+z$-direction, so that the new center is 
$(0,0,1-b)$, with $0 < b < 1$. 
We also assume that the Bloch sphere is deformed into an ellipsoid
with $z$ as symmetry axis:
\begin{equation}
\frac{x^2+y^2}{a^2} + \frac{[z-(1-b)]^2}{b^2}=1.
\label{ellipse}
\end{equation}
Imposing a higher order tangency 
of this ellipsoid to the Bloch sphere
$x^2+y^2+z^2=1$ we obtain $b=a^2$. If we define $a=\cos\theta$
($0<\theta<\pi/2$), then Eq.~(\ref{ellipse}) becomes 
\begin{equation}
\frac{x^2+y^2}{\cos^2\theta} + \frac{(z-\sin^2\theta)^2}
{\cos^4\theta}=1.
\label{ellipse2}
\end{equation}
Note that this equation corresponds to the minimum deformation
required to the Bloch sphere in order to displace its center
along the $z$-axis by $1-b=\sin^2\theta$. 
The graphic visualization 
of the mapping of the Bloch sphere onto an ellipsoid with 
displaced center is shown in Fig.~\ref{fig:ellipse}.

\begin{figure}
\centerline{\epsfxsize=6.5cm\epsffile{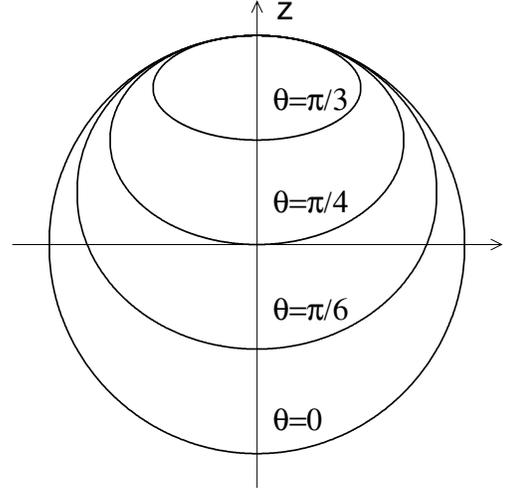}}
\caption{Visualization of the minimum deformation required to
displace the center of the Bloch sphere along the $z$-axis.
The horizontal axis can be any axis in the $(x,y)$ plane.}
\label{fig:ellipse}
\end{figure}

The mapping of the Bloch sphere onto the ellipsoid (\ref{ellipse2})
can be obtained by means of the simple equivalent circuit drawn
in Fig.~\ref{fig:ampdamp}. This circuit leads to a single-qubit
quantum operation known as the amplitude damping channel.
It is described by the Kraus operators
\begin{equation}
F_0=
\left[
\begin{array}{cc}
1 & 0\\ 
0 & \cos\theta
\end{array}
\right],
\quad
F_1=
\left[
\begin{array}{cc}
0 & \sin\theta \\
0 & 0
\end{array}
\right].
\end{equation}
The corresponding transformation of the Bloch sphere coordinates is
\begin{equation}
\left\{
\begin{array}{l}
x^\prime=(\cos \theta) x,\\
y^\prime=(\cos \theta) y,\\
z^\prime=\sin^2 \theta + (\cos^2 \theta) z.
\end{array}
\right.
\end{equation}

\begin{figure}
\centerline{\epsfxsize=8.5cm\epsffile{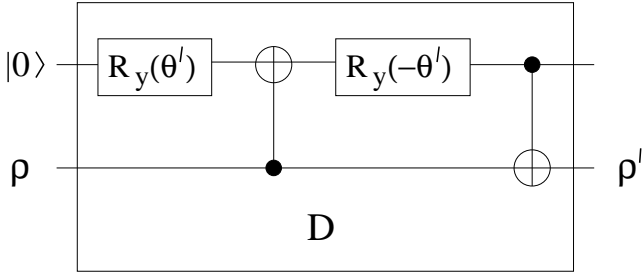}}
\caption{Quantum circuit implementing a displacement of the 
Bloch sphere along the $+z$ direction. Note that $\theta^\prime\equiv
\pi/2-\theta$.}
\label{fig:ampdamp}
\end{figure}

While displacements of the center of the Bloch sphere along 
the positive direction of the $z$-axis can be seen as 
representative of zero temperature dissipation, 
thermal excitations are 
instead associated to displacements along the $-z$-direction.
The equivalent quantum circuit describing thermal excitations
is shown in Fig.~\ref{fig:thermal}. It leads to the Kraus 
operators 

\begin{equation}
F_0=
\left[
\begin{array}{cc}
\cos\theta & 0\\ 
0 & 1
\end{array}
\right],
\quad
F_1=
\left[
\begin{array}{cc}
0 & 0 \\
\sin\theta & 0
\end{array}
\right]
\end{equation}
and to the Bloch sphere coordinate transformation
\begin{equation}
\left\{
\begin{array}{l}
x^\prime=(\cos \theta) x,\\
y^\prime=(\cos \theta) y,\\
z^\prime=-\sin^2 \theta + (\cos^2 \theta) z.
\end{array}
\right.
\end{equation}

\begin{figure}
\centerline{\epsfxsize=8.cm\epsffile{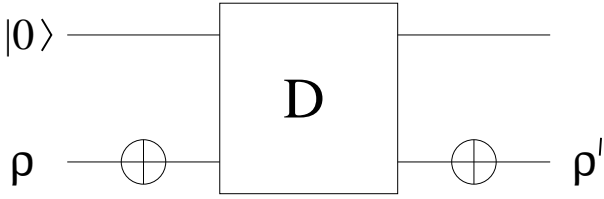}}
\caption{Quantum circuit implementing a displacement of the 
Bloch sphere along the $-z$ direction. The unitary transformation
$D$ corresponds to the boxed part of the circuit in Fig.~\ref{fig:ampdamp}.
The $\oplus$ symbol stands for the NOT gate ($|0\rangle\to |1\rangle,\;
|1\rangle\to|0\rangle$).}
\label{fig:thermal}
\end{figure}

We also consider displacements of the Bloch sphere along
the directions $\pm x$ and $\pm y$. The equivalent 
quantum circuit is drawn in Fig.~\ref{fig:dispxy}.
A displacement along $\pm x$ takes place when the
unitary transformation $U$ in Fig.~\ref{fig:dispxy}
is described by the matrix 
\begin{equation}
U=\frac{1}{\sqrt{2}} 
\left[
\begin{array}{cc}
1 & \pm 1\\ 
\mp 1 & 1
\end{array}
\right].
\end{equation}
The corresponding Kraus operators and the transformation 
of the Bloch sphere coordinates are
\begin{equation}
\begin{array}{c}
F_0=
\frac{1}{2}\left[
\begin{array}{cc}
1+\cos\theta & \pm (1-\cos\theta)\\ 
\pm (1-\cos\theta) & 1+\cos\theta
\end{array}
\right],
\\\\
F_1=
\frac{1}{2}\left[
\begin{array}{cc}
\mp \sin\theta & \sin\theta \\
-\sin\theta & \pm \sin\theta
\end{array}
\right],
\end{array}
\end{equation}
\begin{equation}
\left\{
\begin{array}{l}
x^\prime=\pm \sin^2 \theta + (\cos^2 \theta) x,\\
y^\prime=(\cos \theta) y,\\
z^\prime=(\cos \theta) z.
\end{array}
\right.
\end{equation}
For a displacement along $\pm y$ we have
\begin{equation}
U=\frac{1}{\sqrt{2}} 
\left[
\begin{array}{cc}
1 & \pm i\\ 
\pm i & 1
\end{array}
\right],
\end{equation}
\begin{equation}
\begin{array}{c}
F_0=
\frac{1}{2}\left[
\begin{array}{cc}
1+\cos\theta & \pm i (1-\cos\theta)\\ 
\mp i (1-\cos\theta) & 1+\cos\theta
\end{array}
\right],
\\\\
F_1=
\frac{1}{2}\left[
\begin{array}{cc}
\pm i \sin\theta & \sin\theta \\
\sin\theta & \mp i \sin\theta
\end{array}
\right],
\end{array}
\end{equation}
\begin{equation}
\left\{
\begin{array}{l}
x^\prime=(\cos \theta) x,\\
y^\prime=\pm \sin^2 \theta + (\cos^2 \theta) y,\\
z^\prime=(\cos \theta) z.
\end{array}
\right.
\end{equation}

\begin{figure}
\centerline{\epsfxsize=8.cm\epsffile{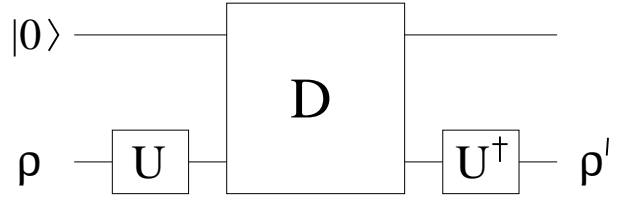}}
\caption{Quantum circuit implementing a displacement of the 
Bloch sphere along the $\pm x$ or $\pm y$ directions. 
The unitary transformation
$D$ corresponds to the boxed part of the circuit in Fig.~\ref{fig:ampdamp}.}
\label{fig:dispxy}
\end{figure}

\end{itemize}

We have given a geometric interpretation of 9 out of the 12
parameters describing a generic single-qubit quantum operation
(3 are associated to rotations about the axes $x$, $y$ or $z$,
3 to displacements along the same axes, and 3 to deformations 
of the Bloch sphere into an ellipsoid, with $x$, $y$ or $z$ as 
symmetry axes). The remaining 3 parameters correspond to 
deformations of the Bloch sphere into an ellipsoid with 
symmetry axis along an arbitrary direction. Since these
deformations can be obtained by combining the 9 previously 
studied quantum operations, then, for small errors, it will be 
sufficient to consider only 9 parameters.

\section{Impact of noise on entanglement purification}
\label{sec:noisyQPA}

We discuss the impact of the 9 noise channels described in 
the previous section on the QPA algorithm.
We present numerical data for the case in which 
quantum noise acts on the top qubit in Fig.~\ref{figDEJMPS}
after the $U$-rotation. However, we point out that 
very similar results are obtained when noise acts on 
one of the other three qubits in the same figure. 
Data are obtained by iteration of a four-qubit noisy 
quantum map, with input state $\rho_{AB}\otimes \rho_{AB}$ 
and output state (for the first two qubits) $\rho_{AB}^\prime$
\cite{footnote}.  

We measure the quality of the purified EPR pair 
by the fidelity $F$, defined in Eq.~(\ref{fido}).
Moreover, we compute the survival probability $P(n)$,
defined in Eq.~(\ref{survp}),
measuring the probability that a $n$-step QPA protocol
is successful. 

We note that the following symmetries in the effect of
errors are observed for the QPA algorithm: 

(i) rotations through an angle $+\theta$ or $-\theta$ have
the same impact;

(ii) displacements along the positive or the negative 
direction of a given axis have the same effect;

(iii) rotations about the $x$ axis and deformations 
with $x$ as symmetry axis (bit flip channel) have 
the same effect; the same observation applies for 
the axes $y$ and $z$ as well. 

The main result of our studies is the demonstration that
the sensitivity of the quantum privacy protocol to errors
strongly depends on the kind of noise. 
Two main distinct behaviors are observed: 

(i) the fidelity is continuously improved by increasing 
the number of purification steps;

(ii) the fidelity saturates to a value $F<1$ after a 
finite number of steps, so that any further iteration
is useless \cite{largeepsilon}.

\begin{figure}
\centerline{\epsfxsize=8.5cm\epsffile{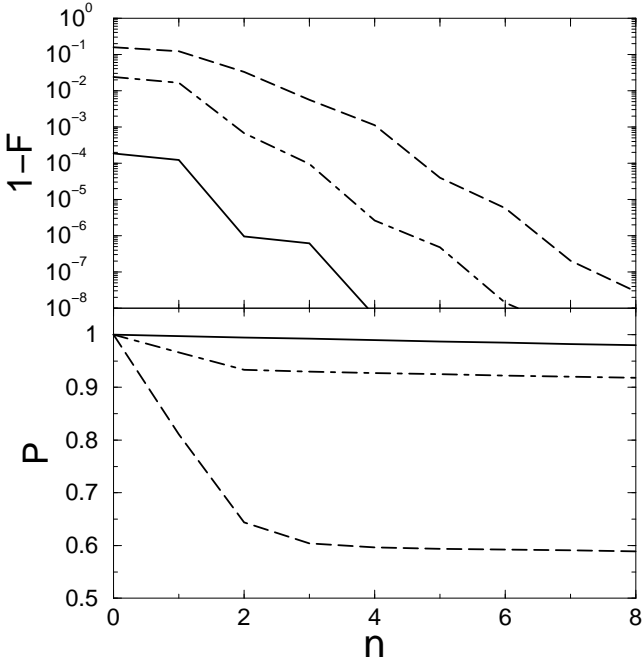}}
\caption{Same as in Fig.~\ref{DEJMPSpure} but 
for the bit flip channel at $\theta=10^{-1}$.} 
\label{fidobitflip}
\end{figure}

\begin{figure}
\centerline{\epsfxsize=8.5cm\epsffile{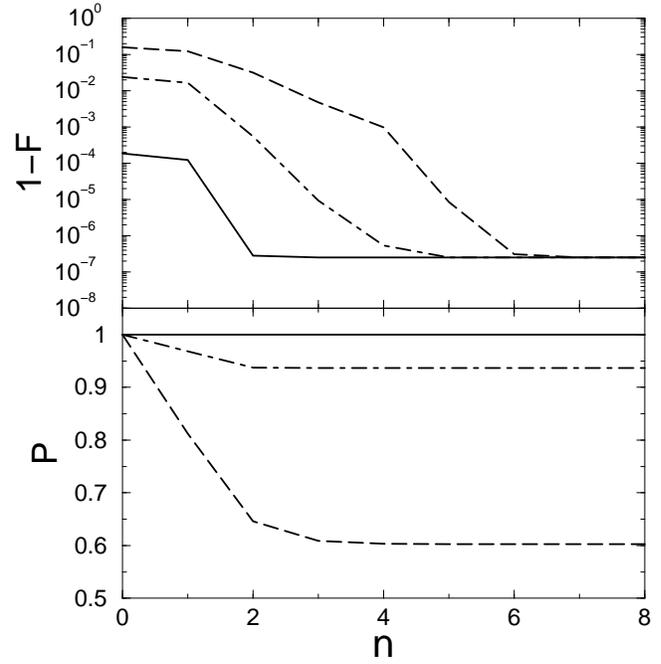}}
\caption{Same as in Fig.~\ref{DEJMPSpure} but for 
the noise channel corresponding to a displacement along
the $x$-axis of the Bloch sphere, at $\theta=10^{-3}$.}
\label{fidodisplacementx}
\end{figure}

As examples of behaviors of the kind (i) and (ii) we show the 
bit-flip channel in Fig.~\ref{fidobitflip} 
(for error strength $\theta=10^{-1}$) and the 
displacement along $x$ in Fig.~\ref{fidodisplacementx}
($\theta=10^{-3}$).
In both figures, the survival probability $P(n)$ can also be seen.
Note that, for these sufficiently small error strengths, 
the values of $P(n)$ shown in Fig.~\ref{fidobitflip} and
Fig.~\ref{fidodisplacementx} are not very far from those of 
the ideal protocol (see Fig.~\ref{DEJMPSpure}).

\begin{figure}
\centerline{\epsfxsize=8.cm\epsffile{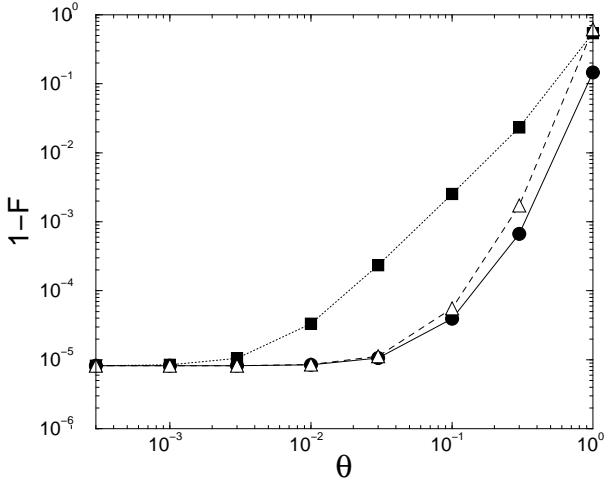}}
\caption{Deviation $1-F$ of the fidelity $F$ from the ideal
case $F=1$ as a function of the noise strength $\theta$,
after $n=5$ steps of the quantum privacy amplification protocol,
for $f_\alpha=0.95$, bit flip (circles), phase flip
(squares) and amplitude damping (triangles) channels.}
\label{DEJMPSnoisy}
\end{figure}

It is important to point out that not only the behavior 
of $F(n)$ is qualitatively different depending on the 
noise channel but also the level of tolerable noise strength 
is channel-dependent.
To give a concrete example, we show in
Fig.~\ref{DEJMPSnoisy} the deviation $1-F$ of the fidelity
from the ideal value $F=1$ as a function of the noise
strength $\theta$. Data are obtained after $n=5$ iterations of
the QPA protocol, in the case of strong Eve's intrusion
($f_\alpha=0.95$) and we consider the bit flip,
the phase flip and the amplitude damping 
(displacement along $z$) channels.
In the noiseless case we start from $1-F=1.57\times 10^{-1}$
and improve the fidelity to $1-F=8.20\times 10^{-6}$ after
$n=5$ iterations of the quantum privacy amplification protocol.
Even though all noise channels degrade the performance
of the protocol, the level of noise that can be safely tolerated
strongly depends on the specific channel.
For instance, it is clear from Fig.~\ref{DEJMPSnoisy} that
the QPA protocol is much more resilient to bit flip and amplitude
damping errors than to phase flip errors.

A further confirmation of the very different impact of the various
noise channels is shown in Table~\ref{tablemaxerr}, showing, at
$f_\alpha=0.95$, the value of $\theta$ such as $1-F=10^{-4}$
after $n=5$ map iterations. This gives an estimate of
the maximum level of error tolerable for each noise channel. 
It is interesting to remark that displacements of the Bloch 
sphere along $x$ and $y$ are much more dangerous than 
displacements along $z$. We note that the value $1-F=10^{-4}$
has been chosen just for convenience but the same conclusions are
obtained also for other values of $1-F$. We also point out that,
as shown in Table~\ref{tablemaxerr}, it is possible to achieve very 
good fidelities in a small number of purification steps also for quite 
high errors $\theta \sim 10^{-1}>>1-F$ affecting the QPA protocol.

\begin{table}
\begin{center}
\begin{tabular}{|l|c|}
\hline
Noise channel & $\theta$ \\
\hline
Rotation about $x$     & $1.55\times 10^{-1}$\\
Rotation about $y$     & $2.69\times 10^{-1}$\\
Rotation about $z$     & $1.92\times 10^{-2}$\\
Bit flip               & $1.55\times 10^{-1}$\\
Bit-phase flip         & $2.69\times 10^{-1}$\\
Phase flip             & $1.92\times 10^{-2}$\\
Displacement along $x$ & $1.91\times 10^{-2}$\\
Displacement along $y$ & $1.91\times 10^{-2}$\\
Displacement along $z$ & $1.27\times 10^{-1}$\\
\hline
\end{tabular}
\end{center}
\caption{Value of the noise strength $\theta$ such that 
$1-F=10^{-4}$ after $n=5$ iterations of the purification 
protocol, at $f_\alpha=0.95$.}
\label{tablemaxerr}
\end{table}

\section{Conclusions} 
\label{sec:conc}

We have performed a systematic study of the effects of the 
different single-qubit noise channels on the quantum privacy
amplification protocol. Our results show the very different 
impact of the various noise channels on the QPA algorithm. 
In particular, we have distinguished between cases where it is 
possible to drive the fidelity arbitrarily close to one and 
others in which the fidelity saturates to a value different
from one. Another important feature that emerges from our
investigations is the strong dependence of the maximum
noise strength tolerable for the QPA protocol on the noise
channel. 
This is a valuable piece of information for experimental 
implementations. 
For instance, the fact that the QPA protocol is much less 
sensitive to displacements along $z$ than along $x$ or $y$
suggests that the $z$-axis is chosen along ``the direction
of noise''. We can then choose the axes $x$ and $y$ to 
minimize other noise effects. Finally, we remark that studies 
like the present one, taking into account all possible single-qubit
quantum noise channels, promise to give useful insights also 
in the field of quantum computation. 

One of us (G.B.) acknowledges support by EU 
(IST-FET EDIQIP contract) and
NSA-ARDA (ARO contract No. DAAD19-02-1-0086).


\begin{thebibliography}{99}

\bibitem{E91}
A.K. Ekert,
Phys. Rev. Lett. {\bf 67}, 661 (1991).

\bibitem{Bennett}
C.H. Bennett, G. Brassard, S. Popescu, B. Schumacher, J.A. Smolin,
and W.K. Wootters, 
Phys. Rev. Lett. {\bf 76}, 722 (1996).

\bibitem{Bennett2}
C.H. Bennett, D.P. DiVincenzo, J.A. Smolin, and W.K. Wootters, 
Phys. Rev. A {\bf 54}, 3824 (1996).

\bibitem{DEJMPS} 
D. Deutsch, A. Ekert, R. Jozsa, C. Macchiavello, S. Popescu, and A. Sanpera,
Phys. Rev. Lett. {\bf 77}, 2818 (1996).

\bibitem{briegel}
H. Aschauer and H.J. Briegel, 
Phys. Rev. A {\bf 66}, 032302 (2002).

\bibitem{buzekhillery}
V. Bu\v{z}ek and M. Hillery,
Phys. Rev. A {\bf 54}, 1844 (1996); quant-ph/9801009.

\bibitem{macchiavello98}
C. Macchiavello,
Phys. Lett. A {\bf 246}, 385 (1998).

\bibitem{chuang}
See, {\it e.g.}, 
M.A. Nielsen and I.L. Chuang,
{\it Quantum computation and quantum information}
(Cambridge University Press, Cambridge, 2000),
Ch. 8.

\bibitem{qcbook}
See, {\it e.g.}, 
G. Benenti, G. Casati, and G. Strini,
{\it Principles of Quantum Computation and Information},
Vol. 1: Basic Concepts (World Scientific, Singapore, 2004),
pag. 110.

\bibitem{footnote}
We have also developed an alternative numerical technique, 
in which $2^n$ imperfect EPR pairs are considered at the beginning,
to obtain, after $n$ steps and when the QPA 
algorithm is successful (coinciding outcomes for Alice and 
Bob's measurements), a single purified EPR pair.
This numerical approach gives the same results as
the simpler four-qubit mapping when
the action of quantum noise is identical for each 
pair of EPR states. However, it has the advantage that 
more general situations can be treated, for instance
the noise strength can depend on the considered pair  
of EPR states.

\bibitem{largeepsilon} 
We note that in the case of strong noise it is also 
possible that the fidelity degrades when the QPA algorithm 
is iterated.

\end{thebibliography}

\appendix
\section{Isotropic cloning}
\label{app:isotropiccloning}

Let us first consider the case in which the initial state 
of Bob's qubit is pure, $|\psi\rangle=\mu|0\rangle+\nu|1\rangle$,
where $\mu,\nu$ are complex numbers, with $|\mu|^2+|\nu|^2=1$.
The unitary transformation $W$ in Fig.~\ref{figBHcrypto} maps 
the state $|\psi\rangle|\Phi\rangle$ (where $|\Phi\rangle$ is 
given by Eq.~(\ref{betgamdel})) onto the state
\begin{equation}
\begin{array}{c}
|\Psi\rangle=
\mu(\alpha|000\rangle+\beta|101\rangle+\gamma|110\rangle
+\delta|011\rangle)
\\
+\nu(\alpha|111\rangle+\beta|010\rangle+\gamma|001\rangle
+\delta|100\rangle).
\end{array}
\end{equation}
We then obtain the density matrix $\rho_B$ after tracing 
the density matrix $|\Psi\rangle\langle\Psi|$ over
Eve's qubit and the ancillary qubit. We have
\begin{equation}
\rho_B=\left[
\begin{array}{cc}
|\mu|^2(\alpha^2+\delta^2) & 2\mu\nu^\star\alpha\delta\\
+|\nu|^2(\beta^2+\gamma^2) & +2\mu^\star\nu\beta\gamma\\
&\\
2\mu^\star\nu\alpha\delta & |\mu|^2(\beta^2+\gamma^2)\\
+2\mu\nu^\star\beta\gamma & +|\nu|^2(\alpha^2+\delta^2)
\end{array}
\right].
\end{equation}
In the same way we obtain the density matrix $\rho_E$ after
tracing over Bob's qubit and the ancillary qubit:
\begin{equation}
\rho_E=\left[
\begin{array}{cc}
|\mu|^2(\alpha^2+\beta^2) & 2\mu\nu^\star\alpha\beta\\
+|\nu|^2(\gamma^2+\delta^2) & +2\mu^\star\nu\gamma\delta\\
&\\
2\mu^\star\nu\alpha\beta & |\mu|^2(\gamma^2+\delta^2)\\
+2\mu\nu^\star\gamma\delta & +|\nu|^2(\alpha^2+\beta^2)
\end{array}
\right].
\end{equation}

Let us call $(x,y,z)$, $(x_B,y_B,z_B)$ and $(x_E,y_E,z_E)$
the Bloch sphere coordinates corresponding to 
$|\psi\rangle\langle\psi|$,
$\rho_B$ and $\rho_E$. We have
\begin{equation}
\mu\nu^\star=\frac{1}{2}(x-iy),\quad
|\mu|^2=\frac{1}{2}(1+z),\quad
|\nu|^2=\frac{1}{2}(1-z).
\end{equation}
After setting $\gamma=0$, we obtain
\begin{equation}
\left\{
\begin{array}{l}
\frac{1}{2}(x_B-iy_B)=(\rho_B)_{01}=(x-iy)\alpha\delta,\\\\
\frac{1}{2}(1+z_B)=(\rho_B)_{00}=
\frac{1}{2}(1+z)(\alpha^2+\delta^2)+\frac{1}{2}(1-z)\beta^2,
\end{array}
\right.
\end{equation}
which imply
\begin{equation}
\left\{
\begin{array}{l}
x_B=2\alpha\delta x,\\
y_B=2\alpha\delta y,\\
z_B=(\alpha^2+\delta^2-\beta^2)z.
\end{array}
\right.
\end{equation}

The state $\rho_B$ is an isotropic cloning of $|\psi\rangle\langle\psi|$
when $R_B=x_B/x=y_B/y=z_B/z$. Therefore we obtain
\begin{equation}
\left\{
\begin{array}{l}
2\alpha\delta=\alpha^2+\delta^2-\beta^2,\\
\alpha^2+\beta^2+\delta^2=1,
\end{array}
\right.
\end{equation}
so that
\begin{equation}
\delta=\frac{\alpha}{2}\pm\sqrt{\frac{1}{2}-\frac{3}{4}\alpha^2}.
\label{deltabh}
\end{equation}
In the same way we obtain
\begin{equation}
\left\{
\begin{array}{l}
x_E=2\alpha\beta x,\\
y_E=2\alpha\beta y,\\
z_E=(\alpha^2+\beta^2-\delta^2)z.
\end{array}
\right.
\end{equation}
Isotropic cloning ($R_E=x_E/x=y_E/y=z_E/z$) is obtained when
\begin{equation}
\left\{
\begin{array}{l}
2\alpha\beta=\alpha^2+\beta^2-\delta^2,\\
\alpha^2+\beta^2+\delta^2=1,
\end{array}
\right.
\end{equation}
so that
\begin{equation}
\beta=\frac{\alpha}{2}\pm\sqrt{\frac{1}{2}-\frac{3}{4}\alpha^2}.
\label{betabh}
\end{equation}
Note that, if we choose the plus sign in (\ref{deltabh}), then 
the minus sign has to be taken in (\ref{betabh}) in order that
the normalization condition $\alpha^2+\beta^2+\delta^2$ is satisfied.
This choice corresponds to Eq.~(\ref{betadeltabh}).

Note that the cloning is isotropic also in the case in which the
initial state $\rho$ of Bob's qubits is mixed. In this case we 
can write $\rho=\sum_i p_i \rho_i$, with 
$\rho_i=|\psi_i\rangle\langle\psi_i|$ pure state. 
The Bloch vector ${\bf r}$ associated to $\rho$ is the weighted
sum of the Bloch vectors ${\bf r}_i$ associated to the density matrices
$\rho_i$: ${\bf r}=\sum_i p_i {\bf r}_i$. Since we have seen that for 
pure initial states
$({\bf r}_i)_B= R_B {\bf r}_i$ and
$({\bf r}_i)_E= R_E {\bf r}_i$, then
${\bf r}_B=\sum_i p_i ({\bf r}_i)_B= R_B {\bf r}$ and
${\bf r}_E=\sum_i p_i ({\bf r}_i)_E= R_E {\bf r}$.

\end{document}